\documentclass[12pt]{article}
\usepackage{amsmath}
\usepackage{graphicx}
\usepackage{subfigure}
\usepackage{hyperref}

\begin{document}

\author{Ernst Trojan \and \textit{Moscow Institute of Physics and Technology} \and 
\textit{PO Box 3, Moscow, 125080, Russia}}
\title{Analytic theory of discontinuities in current-carrying cosmic strings}
\maketitle

\begin{abstract}
We formulate an analytic method to study the discontinuities in
superconducting cosmic strings. Equations of discontinuities and conditions
of their existence are derived from the intrinsic and extrinsic equations of
motion. It is the fundamental for research of particular solutions,
associated with kinks, cusps and shocks.
\end{abstract}

\section{Introduction}

Cosmic strings are 1+1-dimensional topological defects which are believed to
be formed in the era of cosmological phase transitions and which are
responsible for the global structure of the Universe \cite{HK}. Although
their experimental recognition is not evident, the physical presence of
cosmic strings is reflected in several astrophysical phenomena associated
with gravitational lensing, gravitational waves, particle acceleration,
cosmic microwave background and gamma-ray bursts.

When the curvature of a string is small with respect to its characteristic
width (core), the surface Lagrangian $\Lambda $\ is determined through
integration over extrinsic coordinates. The the starting point is the
Goto-Nambu model $\Lambda =-m^2$, and its extension is required when a
cosmic string acquires non-trivial internal structure, associated with
actual particles (bosons or fermions) which are trapped in the defect core 
\cite{W85}. The Lagrangian of this string $\Lambda =\Lambda (\chi )$ becomes
dependent on the magnitude of the current $\chi =-\partial ^a\psi \partial
_a\psi $ where the gradient is taken over the worldsheet coordinates and the
phase $\psi $ originates from the wave function of the field, condensed in
the string core.

The equations of motion of such current-carrying or ''superconducting''
string include a pair of ''intrinsic'' equations \cite{C89a} 
\begin{equation}
\eta _\mu ^\nu \nabla _\nu \,\left( \mu v^\mu \right) =0  \label{cur1}
\end{equation}
\begin{equation}
\eta _\mu ^\nu \nabla _\nu \,\left( nu^\mu \right) =0  \label{cur2}
\end{equation}
and a pair of ''extrinsic'' equations 
\begin{equation}
\perp _\rho ^\mu \left( u^\nu \nabla _\nu v^\rho -v^\nu \nabla _\nu u^\rho
\right) =0  \label{ex1}
\end{equation}
\begin{equation}
\perp _\rho ^\mu \left( Uu^\nu \nabla _\nu u^\rho -Tv^\nu \nabla _\nu v^\rho
\right) =0  \label{ex2}
\end{equation}
where 
\begin{equation}
\eta _\mu ^\nu =v^\nu v_\mu -u^\nu u_\mu \qquad \perp _\mu ^\nu =g_\mu ^\nu
-\eta _\mu ^\nu  \label{pro}
\end{equation}
are the parallel and orthogonal projective tensors, composed of mutually
orthogonal time-like and space-like unit vectors 
\begin{equation}
u^\mu u_\mu =-v^\mu v_\mu =-1\qquad u^\mu v_\mu =0  \label{un}
\end{equation}
and quantities $U$, $T$, $\mu =dU/dn$, $n=-dT/d\mu $ are dependent on the
current $\chi $ and obey relations $\mu ^2=K^2n^2=\bar K^{-2}n^2$ at
space-like currents (''magnetic'' regime) and $\mu ^2=K^{-2}n^2=\bar K^2n^2$
at time-like currents (''electric'' regime) where parameter $K=\bar
K^{-1}=2d\Lambda /d\chi $ is determined by explicit functional dependence $%
\Lambda \left( \chi \right) $ or $\mu \left( n\right) $, called as equation
of state (EOS) \cite{CP95}.

The ''intrinsic'' equations (\ref{cur1})-(\ref{cur2}) admit infinitesimal
perturbations within the worldsheet, which are similar to longitudinal
waves, propagating at the speed \cite{C89a} 
\begin{equation}
c_L^2=-\frac{dT}{dU}=\frac n\mu \frac{d\mu }{dn}  \label{cc}
\end{equation}
The ''extrinsic'' equations (\ref{ex1})-(\ref{ex2}) admit infinitesimal
perturbations of the string worldsheet , which are similar to transversal
waves, propagating at the speed 
\begin{equation}
c_E^2=\frac TU  \label{bb}
\end{equation}

Finite-amplitude perturbations of the string worldsheet are known as kinks
and cusps \cite{C1,C2,C3,C4}. A kink is known as abrupt changes of the
curvature of the string $\kappa $. A cusp is another type of geometric
phenomenon when a part of the string is inflected and doubled on itself,
moving at the speed of light. Most research of these perturbations is
devoted to the Goto-Nambu strings. The current-carrying strings may admit
not only kinks and cusps but a principally new class of perturbations called
as shocks, predicted earlier \cite{V99}, investigated numerically \cite
{MP00,CCMP02} and explained analytically \cite{TV2012}. In contrast to the
kinks and cusps, the shocks do not deal with the string geometry and occur
within the worldsheet. A shock originates from a longitudinal (or ''sound'')
wave (\ref{cc}) when its amplitude grows up and the current $\chi $ becomes
discontinuous that can be described by explicit formulas \cite{TV2012}. The
kink formation in superconducting string is studied numerically \cite
{MP00,CCMP02} but no analytic formula has been proposed for explanation of
its physical nature.

The finite-amplitude perturbations may play important role in the evolution
of closed string configurations (vortons) and string networks, they can be
responsible for various observable effects \cite{C44,C5,C6,C7}. The
structures, triggered by finite-amplitude variations of the current, may
also appear during reconnection or self-intersection of the string loops,
when the current increases beyond the range of the vorton stability, that in
all cases will result in visible radiation events. Although a string with
finite current ($\chi \neq 0$) may reveal qualitatively new properties,
which are absent in the chiral case ($\chi =0$), a relationship between
perturbations of the current and the geometry can be scarcely known without
numerical computation. It is still unknown whether the kink velocity depends
on the jump of the current $\Delta \chi $, how it is dependent on the jump
of the curvature $\Delta \kappa $, can a shock produce changes in the
curvature, which perturbations of the curvature are possible besides the
kinks, and which perturbations of the current are possible besides the
shocks?

Indeed, a universal method at the analytic level is highly desirable but it
seems to be extremely complicated, almost like the numerical solution.
Although the kinks, cusps and shocks are no more than discontinuous
solutions of the equations of motion (\ref{cur1})-(\ref{ex2}), and the
relevant mathematical theory is developed \cite{A1989} but it is not easy to
adjust it to the string equations of motion. It is the main difficulty and
it is the purpose of the present paper. We derive the basic equations of
discontinuities in the current-carrying cosmic strings and outline several
ideas for their further analysis. Particular solutions and discussions will
be given in separate study.

\section{Discontinuities in differential equations}

The general theory of discontinuous solution of ordinary differential
equations and its applications are well known \cite{A1989}. In application
to our problem, the front of a discontinuity is a hypersurface in
4-dimensional space whose equation is given by a scalar function $\zeta
\left( x\right) =0$, which acquires distinct values $\zeta _{+}\neq \zeta
_{-}$ for the states before and behind the front, labeled by ''$-$'' and ''$%
+ $''. As a result, an arbitrary tensor distribution $\Omega =\Omega _{\beta
...}^{\alpha ...}$ may reveal a jump $\Omega _{+}\equiv \Omega \left(
x,\zeta _{+}\right) \neq \Omega _{-}\equiv \Omega \left( x,\zeta _{-}\right) 
$, which depends on the direction of the unit space-like characteristic
vector 
\begin{equation}
\lambda _\mu =\frac{d\zeta }{dx^\mu }\qquad \lambda ^\mu \lambda _\mu =1
\label{lam}
\end{equation}
A numerical analysis of the equations of motion can be applied to the
process of birth, growth (or decay) and stable phase of discontinuities. The
analytic analysis, developed in the present paper, is applied to the stable
discontinuities, characterized by a stable front with constant
characteristic vector 
\begin{equation}
\lambda _\mu =\lambda _{+\mu }=\lambda _{-\mu }  \label{stab}
\end{equation}

Appearance of a discontinuity implies that the covariant derivative of an
arbitrary tensor $\Omega \left( x,\zeta \right) $ is presented in the
extended form \cite{A1989} 
\begin{equation}
\nabla _\nu \Omega _{+}=\nabla _\nu \Omega _{-}+\lambda _\nu \left[
D_\lambda \Omega \right]  \label{div0}
\end{equation}
where differentiation along the characteristic direction is included in
square brackets: 
\begin{equation}
\left[ D_\lambda \Omega \right] \equiv \lambda ^\mu \left[ \nabla _\mu
\Omega \right] =\lambda ^\mu \frac{\partial \Omega }{\partial \zeta }\frac{%
d\zeta }{dx^\mu }=\frac{d\Omega }{d\zeta }  \label{ln}
\end{equation}
We expect no gravitational wave and consider the same background metric $%
g_{\mu \nu }=g_{+\mu \nu }=g_{-\mu \nu }$ at both sides of the
discontinuity. Coming to finite increment $d\zeta \rightarrow \zeta
_{+}-\zeta _{-}$, in the frames of linear approximation, we write 
\begin{equation}
\lambda _\nu \left[ D_\lambda \Omega \right] =\lambda _\nu \frac{\Omega
_{+}-\Omega _{-}}{\zeta _{+}-\zeta _{-}}  \label{ln0}
\end{equation}
If tensor $\Omega $ obeys the equation of motion 
\begin{equation}
\nabla _\nu \Omega =0  \label{emo0}
\end{equation}
formula (\ref{div0}) implies $\lambda _\nu \left[ D_\lambda \Omega \right]
=0 $. Operating with arbitrary $\Delta \zeta =\zeta _{+}-\zeta _{-}$ in (\ref
{ln0}), we derive the equation of discontinuity 
\begin{equation}
\lambda _\nu \left( \Omega _{+}-\Omega _{-}\right) =0  \label{dis0}
\end{equation}
It states that variables $\Omega _{+}$ and $\Omega _{-}$ may differ but
their projection unto the characteristic direction $\lambda _\nu $ must
coincide at both sides of the discontinuity. The acoustic limit $\Omega
_{+}\rightarrow \Omega _{-}$ corresponds to identical states before and
behind the front when the discontinuity decays.

Tensor $\Omega $ may obey a more sophisticated equation of motion 
\begin{equation}
\Xi \nabla _\nu \Omega =0  \label{emo1}
\end{equation}
with dual tensor $\Xi =\Xi _{\sigma ...}^{\rho ...}$. Substituting formula (%
\ref{div0}) in equation (\ref{emo1}), we have 
\begin{equation}
0=\Xi _{+}\nabla _\nu \Omega _{+}=\Xi _{+}\nabla _\nu \Omega _{-}+\lambda
_\nu \Xi _{+}\left[ D_\lambda \Omega \right]  \label{div1}
\end{equation}
\begin{equation}
0=\Xi _{-}\nabla _\nu \Omega _{-}=\Xi _{-}\nabla _\nu \Omega _{+}-\lambda
_\nu \Xi _{-}\left[ D_\lambda \Omega \right]  \label{div11}
\end{equation}
The terms in square brackets in the right side of (\ref{div1})-(\ref{div11})
yield the equation of discontinuity 
\begin{equation}
\lambda _\nu \Xi _{+}\left( \Omega _{+}-\Omega _{-}\right) =\lambda _\nu \Xi
_{-}\left( \Omega _{+}-\Omega _{-}\right) =0  \label{dis1}
\end{equation}
under additional condition 
\begin{equation}
\Xi _{+}\nabla _\nu \Omega _{-}=\Xi _{-}\nabla _\nu \Omega _{+}=0
\label{coc1}
\end{equation}
If $\Xi $ is a vector or a scalar, constraint (\ref{coc1}) provides the
ultimate possibility $\Xi _{+}=C\Xi _{-}$ with scalar constant $C\neq 0$. In
general, we must test each solution of equation (\ref{dis1}) whether it
satisfies requirement (\ref{coc1}).

We may also deal with a multi-component system, which includes a set of
tensors $\Omega _i$ (and $\Xi _i$), and the equation of motion in the form

\begin{equation}
\Xi _i\nabla _\nu \Omega _i=0  \label{emo2}
\end{equation}
where summation over index $i$ is performed. Equations (\ref{dis1})-(\ref
{coc1}), then, are written so 
\begin{equation}
\lambda _\nu \Xi _{+i}\left( \Omega _{+i}-\Omega _{-i}\right) =0\qquad
\lambda _\nu \Xi _{-i}\left( \Omega _{+i}-\Omega _{-i}\right) =0
\label{dis2}
\end{equation}
and 
\begin{equation}
\Xi _{+i}\nabla _\nu \Omega _{-i}=\Xi _{-i}\nabla _\nu \Omega _{+i}=0
\label{coc2}
\end{equation}

After all, we should bear in mind an important principle of resolvability 
\cite{LL87}. When we are looking for solution of differential equations in
the form of discontinuity, any initial perturbation is defined by some
number of free parameters. Its evolution is governed by a set of boundary
conditions, corresponding to the equations of motions. A perturbation can
exist in a stable form if the number of these equations coincides with the
number of free parameters in them. If there are too many parameters, the
solution is undefined; if there are only few parameters, the system is
unresolved. There may also exist a degenerate solution which automatically
satisfies all the equations without regard of the free parameters.

\section{Discontinuities in cosmic strings}

Let us apply the common ideology of discontinuities (\ref{emo1})-(\ref{coc2}%
) to the intrinsic and extrinsic equations of motion of a cosmic string.
Applying formula (\ref{div1}) to equation (\ref{cur1}), we have 
\begin{equation}
\eta _{\mu +}^\nu \nabla _\nu \left( n_{+}u_{+}^\mu \right) =\eta _{\mu
+}^\nu \nabla _\nu \left( n_{-}u_{-}^\mu \right) +\lambda _\nu \eta _{\mu
+}^\nu \left[ D_\lambda \left( nu^\mu \right) \right]  \label{in0}
\end{equation}
so that the equation of discontinuity (\ref{dis1}) is written in the form 
\begin{equation}
\lambda _\nu \eta _{+\mu }^\nu \left( n_{+}u_{+}^\mu -n_{-}u_{-}^\mu \right)
=\lambda _\nu \eta _{-\mu }^\nu \left( n_{+}u_{+}^\mu -n_{-}u_{+}^\mu
\right) =0  \label{in1}
\end{equation}
and it is valid under condition

\begin{equation}
\eta _{\mu +}^\nu \nabla _\nu \left( n_{-}u_{-}^\mu \right) =\eta _{\mu
-}^\nu \nabla _\nu \left( n_{+}u_{+}^\mu \right) =0  \label{in11}
\end{equation}

Applying formula (\ref{div1}) to intrinsic equation of motion (\ref{cur2}),
we have 
\begin{equation}
\eta _{\mu +}^\nu \nabla _\nu \left( \mu _{+}v_{+}^\mu \right) =\eta _{\mu
+}^\nu \nabla _\nu \left( \mu _{-}v_{-}^\mu \right) +\lambda _\nu \eta _{\mu
+}^\nu \left[ D_\lambda \left( \mu v^\mu \right) \right]  \label{so0}
\end{equation}
that results in the equation of discontinuity 
\begin{equation}
\lambda _\nu \eta _{+\mu }^\nu \left( \mu _{+}v_{+}^\mu -\mu _{-}v_{-}^\mu
\right) =0\qquad \lambda _\nu \eta _{-\mu }^\nu \left( \mu _{+}v_{+}^\mu
-\mu _{-}v_{-}^\mu \right)  \label{in2}
\end{equation}
which is valid under condition 
\begin{equation}
\eta _{\mu +}^\nu \nabla _\nu \left( \mu _{-}v_{-}^\mu \right) =\eta _{\mu
-}^\nu \nabla _\nu \left( \mu _{+}v_{+}^\mu \right) =0  \label{in22}
\end{equation}

Discontinuities in extrinsic equations are determined by the same method.
Defining expressions 
\begin{equation}
\Xi _1=\perp _\rho ^\mu v^\nu \qquad \Omega _1=u^\rho \qquad \Xi _2=-\perp
_\rho ^\mu u^\nu \qquad \Omega _2=v^\rho  \label{xx1}
\end{equation}
we put them in equation (\ref{ex1}) and, applying formula (\ref{dis2}), we
obtain equations 
\begin{equation}
\lambda _\nu \perp _{+\rho }^\mu \left( v_{+}^\nu u_{-}^\rho -u_{+}^\nu
v_{-}^\rho \right) =0  \label{lu1}
\end{equation}
\begin{equation}
\lambda _\nu \perp _{-\rho }^\mu \left( v_{-}^\nu u_{+}^\rho -u_{-}^\nu
v_{+}^\rho \right) =0  \label{lu2}
\end{equation}
which are valid under condition (\ref{coc1}) that is 
\begin{equation}
\perp _{+\rho }^\mu \left( v_{+}^\nu \nabla _\nu u_{-}^\rho -u_{+}^\nu
\nabla _\nu v_{-}^\rho \right) =0\qquad \perp _{-\rho }^\mu \left( v_{-}^\nu
\nabla _\nu u_{+}^\rho -u_{-}^\nu \nabla _\nu v_{+}^\rho \right) =0
\label{cc1}
\end{equation}

Defining expressions 
\begin{equation}
\tilde \Xi _1=U\perp _\rho ^\mu u^\nu \qquad \tilde \Omega _1=u^\rho \qquad
\tilde \Xi _2=-T\perp _\rho ^\mu v^\nu \qquad \tilde \Omega _2=v^\rho
\label{xx2}
\end{equation}
we put then in equation (\ref{ex2}) and, applying formula (\ref{dis2}), we
obtain equations 
\begin{equation}
\lambda _\nu \perp _{+\rho }^\mu \left( u_{+}^\nu u_{-}^\rho
-c_{E+}^2v_{+}^\nu v_{-}^\rho \right) =0  \label{lu3}
\end{equation}
\begin{equation}
\lambda _\nu \perp _{-\rho }^\mu \left( u_{-}^\nu u_{+}^\rho
-c_{E-}^2v_{-}^\nu v_{+}^\rho \right) =0  \label{lu4}
\end{equation}
which are valid under condition (\ref{coc2}) that is

\begin{equation}
\perp _{+\rho }^\mu \left( u_{+}^\nu \nabla _\nu u_{-}^\rho
-c_{E+}^2v_{+}^\nu \nabla _\nu v_{-}^\rho \right) =0\qquad \perp _{-\rho
}^\mu \left( u_{-}^\nu \nabla _\nu u_{+}^\rho -c_{E-}^2v_{-}^\nu \nabla _\nu
v_{+}^\rho \right) =0  \label{cc2}
\end{equation}

\section{Characteristic vector}

We are still unable to solve the equations until we include an explicit
expression of the space-like characteristic vector (\ref{lam}) and a link
between the physical parameters before and behind the front. Let us present
the characteristic vector as a sum 
\begin{equation}
\lambda _{\pm }^\mu =\alpha _{\pm }^\mu +\sigma _{\pm }^\mu   \label{la12}
\end{equation}
of longitudinal component 
\begin{equation}
\alpha _{\pm }^\mu =\eta _{\pm \nu }^\mu \lambda _{\pm }^\nu =\eta _{\pm \nu
}^\mu \alpha _{\pm }^\nu   \label{al}
\end{equation}
and transversal component 
\begin{equation}
\sigma _{\pm }^\mu =\perp _{\pm \nu }^\mu \lambda _{\pm }^\nu =\perp _{\pm
\nu }^\mu \sigma _{\pm }^\nu   \label{be}
\end{equation}
where 
\begin{equation}
\alpha _{\pm }^2\equiv \alpha _{\pm }^\mu \alpha _{\pm \mu }=1-\sigma _{\pm
}^\mu \sigma _{\pm \mu }\equiv 1-\sigma _{\pm }^2\leq 1  \label{alpha}
\end{equation}
The components (\ref{al}) and (\ref{be}) are mutual orthogonal $\alpha
_{+}^\mu \sigma _{+\mu }=\alpha _{-}^\mu \sigma _{-\mu }=0$ because 
\begin{equation}
\perp _{\pm \mu }^\nu \alpha _{\pm \nu }=\eta _{\pm \mu }^\nu \sigma _{\pm
\nu }=0\qquad \eta _{\pm \nu }^\mu \perp _{\pm \mu }^\rho =0  \label{etab0}
\end{equation}
according to the properties of projective tensor (\ref{pro}). It is also
clear that the transversal component (\ref{be}) is orthogonal to each
worldsheet vector $\sigma _{\pm }^\mu u_{\pm \mu }=\sigma _{\pm }^\mu v_{\pm
\mu }=0$.  

The space-like longitudinal component (\ref{al}) can be presented as a
linear combination of the worldsheet vectors 
\begin{equation}
\alpha _{\pm }^\mu =\alpha _{\pm }\frac{w_{\pm }u_{\pm }^\mu +v_{\pm }^\mu }{%
\sqrt{1-w_{\pm }^2}}  \label{alf}
\end{equation}
so that formula (\ref{al}) implies 
\begin{equation}
\lambda _{\pm \nu }u_{\pm }^\nu =-\frac{\alpha _{\pm }w_{\pm }}{\sqrt{%
1-w_{\pm }^2}}\qquad \lambda _{\pm \nu }v_{\pm }^\nu =\frac{\alpha _{\pm }}{%
\sqrt{1-w_{\pm }^2}}  \label{luv}
\end{equation}
where $w_{-}$\ and $w_{+}$ are the velocities before and behind the front,
in the preferred reference frame, co-moving the discontinuity. When we
operate in the laboratory reference frame, the front of the discontinuity
propagates at velocity $W$ with respect to the string and there is no motion
before the discontinuity $W_{-}=0$ but we do not exclude a possibility of
motion behind the front $W_{+}\neq 0$ (as a result of the discontinuity).
When we switch to the co-moving reference frame, the front is at rest ($%
W-W\equiv 0$), and we operate with finite flow before the front $w_{-}=-W$
and finite flow behind the front $w_{+}=\left( W_{+}-W\right) /\left(
1-Ww_{+}\right) $. Particularly, $W_{+}=W_{-}=0$ corresponds to $w_{+}=w_{-}$%
.

\section{Equations of extrinsic and intrinsic discontinuities}

Consider the equations of extrinsic discontinuities (\ref{lu1}), (\ref{lu2}%
), (\ref{lu3}) and (\ref{lu4}). By means of (\ref{pro}) and (\ref{gh}) we
get expressions
\begin{equation}
\perp _{+\rho }^\mu v_{-}^\rho =v_{-}^\mu -Xv_{+}^\mu +Gu_{+}^\mu \qquad
\perp _{+\rho }^\mu u_{-}^\rho =u_{-}^\mu -Hv_{+}^\mu +Yu_{+}^\mu 
\label{ort1}
\end{equation}
\ 
\begin{equation}
\perp _{-\rho }^\mu v_{+}^\rho =v_{+}^\mu -Xv_{-}^\mu +Hu_{-}^\mu \qquad
\perp _{-\rho }^\mu u_{+}^\rho =u_{+}^\mu -Gv_{-}^\mu +Yu_{-}^\mu 
\label{ort2}
\end{equation}
where

\begin{equation}
G=v_{-}^\mu u_{+\mu }\qquad H=u_{-}^\mu v_{+\mu }\qquad X=v_{-}^\mu v_{+\mu
}\qquad Y=u_{-}^\mu u_{+\mu }  \label{gh}
\end{equation}
Substituting formulas (\ref{luv}) and (\ref{ort1})-(\ref{ort2}) in equations
(\ref{lu1})-(\ref{lu2}) and (\ref{lu3})-(\ref{lu4}), we have 
\begin{equation}
\alpha _{+}\left\{ u_{-}^\mu -Hv_{+}^\mu +Yu_{+}^\mu +w_{+}\left( v_{-}^\mu
-Xv_{+}^\mu +Gu_{+}^\mu \right) \right\} =0  \label{lu11}
\end{equation}
\begin{equation}
\alpha _{-}\left\{ u_{+}^\mu -Gv_{-}^\mu +Yu_{-}^\mu +w_{-}\left( v_{+}^\mu
-Xv_{-}^\mu +Hu_{-}^\mu \right) \right\} =0  \label{lu22}
\end{equation}
\begin{equation}
\alpha _{+}\left\{ -w_{+}\left( u_{-}^\mu -Hv_{+}^\mu +Yu_{+}^\mu \right)
-c_{E+}^2\left( v_{-}^\mu -Xv_{+}^\mu +Gu_{+}^\mu \right) \right\} =0
\label{lu33}
\end{equation}
\begin{equation}
\alpha _{-}\left\{ -w_{-}\left( u_{+}^\mu -Gv_{-}^\mu +Yu_{-}^\mu \right)
-c_{E-}^2\left( v_{+}^\mu -Xv_{-}^\mu +Hu_{-}^\mu \right) \right\} =0
\label{lu44}
\end{equation}

Equations of intrinsic discontinuities are obtained by substituting formula (%
\ref{al}) in equations (\ref{in1})-(\ref{in2}), namely 
\begin{equation}
\alpha _{+\mu }\left( n_{+}u_{+}^\mu -n_{-}u_{-}^\mu \right) =0\qquad \alpha
_{-\mu }\left( n_{+}u_{+}^\mu -n_{-}u_{+}^\mu \right) =0  \label{so11}
\end{equation}
\begin{equation}
\alpha _{+\mu }\left( \mu _{+}v_{+}^\mu -\mu _{-}v_{-}^\mu \right) =0\qquad
\alpha _{-\mu }\left( \mu _{+}v_{+}^\mu -\mu _{-}v_{-}^\mu \right)
\label{so22}
\end{equation}
Taking into account definition (\ref{alf}) and coefficients (\ref{gh}), we
immediately obtain

\begin{equation}
-n_{+}\alpha _{+}w_{+}=n_{-}\alpha _{+}\left( H+Yw_{+}\right) \qquad
n_{+}\alpha _{-}\left( G+Yw_{-}\right) =-n_{-}\alpha _{-}w_{-}  \label{so1}
\end{equation}
\begin{equation}
\mu _{+}\alpha _{+}=\mu _{-}\alpha _{+}\left( X+Gw_{+}\right) \qquad \mu
_{+}\alpha _{-}\left( X+Hw_{-}\right) =\mu _{-}\alpha _{-}  \label{so2}
\end{equation}

\section{Classes of discontinuities}

As we have mentioned above, existence of a stable discontinuous solution is
possible when the number of equations of motion coincide with the number of
free parameters. A cosmic string is described by four equations of motion.
However, the total system of four equations is split into the intrinsic pair
(\ref{cur1})-(\ref{cur2}) and extrinsic pair (\ref{ex1})-(\ref{ex2}), where
intrinsic and extrinsic motions are considered separately \cite
{C89a,MP00,CCMP02}. Therefore, we may expect a 4-parametric perturbation,
which is decomposed in an sum of intrinsic and extrinsic discontinuities. An
extrinsic discontinuity is described by 2 extrinsic equations with 2
parameters. An intrinsic discontinuity is described by 2 intrinsic equations
with 2 parameters. 

The first free parameter is the velocity of the discontinuity $W$ which is
determined by the physical state of the string. The second parameter is the
increment of the current $\Delta \chi $ or the increment of the curvature $%
\Delta \kappa $ (but simultaneous change of the current and the curvature is
not admitted in a 2-parametric discontinuity). The previous analytical
analysis of infinitesimal perturbations \cite{C89a} have revealed existence
of intrinsic perturbations of the current and extrinsic perturbations of the
curvature. Intrinsic perturbations of the curvature and extrinsic
perturbations of the current do not exist. We cannot expect their appearance
at finite amplitude of the perturbation because this threshold amplitude
becomes the 3rd parameter which is forbidden. \cite{MP00,CCMP02}

If we suppose that intrinsic and extrinsic perturbations can form a
composite discontinuity it will be a 3-parametric discontinuity in 4-folded
system of equation that is also forbidden. Therefore, intrinsic and
extrinsic discontinuities cannot co-exist in the same solution. 

As for the degenerate discontinuity, the evident trivial solution 
\begin{equation}
\alpha _{+}=\alpha _{-}=0  \label{aa00}
\end{equation}
satisfies automatically both the extrinsic equations (\ref{lu11})-(\ref{lu44}%
) and the intrinsic equations (\ref{so1})-(\ref{so2}). No change of the
curvature and no change of the current is incorporated in this discontinuity
because all these parameters remain undetermined since the equations are
satisfied automatically. A degenerate discontinuity must propagate at
constant velocity (independent on the physical state of the string and its
curvature) that can correspond to zero or to the speed of light. Note that
equality (\ref{aa00}) implies that the characteristic vector (\ref{be}) is
orthogonal to the string worldsheet. We shall see in the next study that
these degenerate discontinuities are no more than the \textit{cusps}.

\section{Conclusion}

The general form of equation of motion (\ref{emo1}) can admit discontinuous
solution (\ref{dis1}) under condition (\ref{coc1}). If equation of motion is
given in a multi-component form (\ref{emo1}), the equations of
discontinuities (\ref{dis2}) are solved under condition (\ref{coc2}). 

Superconducting cosmic strings can admit two classes of discontinuities. The
intrinsic discontinuities of the current are determined by equations (\ref
{in1}), (\ref{in2}) under constraints (\ref{in11}), (\ref{in22}). The
extrinsic discontinuities of the curvature are determined by equations (\ref
{lu1}), (\ref{lu2}), (\ref{lu3}), (\ref{lu4}) under constraints (\ref{cc1}),
(\ref{cc2}). Having introduced the characteristic vector (\ref{la12}), (\ref
{alf}), we establish equations of extrinsic discontinuities (\ref{lu11})-(%
\ref{lu44}) and equations of intrinsic discontinuities (\ref{so1})-(\ref{so2}%
). It is the main result of the present paper. 

There is also a degenerate solution (\ref{aa00}) which satisfies all the
equations of discontinuities and does not concern any change of the current
or the curvature.

Looking for explicit solution of equations (\ref{lu11})-(\ref{lu44}) and (%
\ref{so1})-(\ref{so2}), we should utilize the important condition of
stability (\ref{stab}) of the characteristic vector (\ref{la12}), (\ref{alf}%
). Explicit results will be obtained in the next papers. 


\end{document}